\documentclass[a4paper,11pt]{article}

\usepackage[latin1]{inputenc}

\usepackage[T1]{fontenc}
\usepackage[dvips]{graphicx}
\usepackage{times}

\usepackage{pstricks}
\usepackage{calrsfs}
\usepackage{subeqn}

\usepackage{pifont}
\usepackage{oldgerm}
\usepackage{color} 
\usepackage{amssymb,amsbsy}
\usepackage{epic,eepic}
\usepackage{texdraw}
\usepackage{dsfont}
\usepackage{latexsym}

\usepackage{accents}

\newtheorem{The}{Theorem}

\newtheorem{Cor}{Corollary}

\newcommand{\nn}{\nonumber}

\begin{document}

\title{Symmetries and conservation laws of lattice Boussinesq equations}
\author{Pavlos Xenitidis and Frank Nijhoff \\ 
School of Mathematics, University of Leeds, LS2 9JT, Leeds, UK \\
P.Xenitidis@leeds.ac.uk, F.W.Nijhoff@leeds.ac.uk}
\date{}
\maketitle

\begin{abstract}
Sequences of canonical conservation laws and generalized symmetries for the lattice  Boussinesq and the lattice modified Boussinesq systems are successively 
derived. The interpretation of these symmetries as differential-difference equations leads to corresponding hierarchies of such equations for which conservation 
laws and Lax pairs are constructed. Finally, using the continuous symmetry reduction approach, an integrable, multidimensionally consistent system of partial 
differential equations is derived in relation with the lattice modified Boussinesq system.
\end{abstract}

\section{Introduction}

The existence of infinite hierarchies of symmetries and conservation laws is among the key characteristics of most of the well known integrable systems, 
\cite{M,O,Z}. Recently, in \cite{MWX1}, this characteristic of integrability was also established for a class of two-dimensional integrable partial difference equations, 
namely the equations in the Adler-Bobenko-Suris (ABS) list, cf. \cite{ABS}, of canonical scalar affine-linear lattice equations. Moreover, in \cite{X}, it was shown that equations 
in this class possess infinite hierarchies of symmetries and conservation laws, and these hierarchies were constructed recursively. Their derivation is similar to the 
approach used in \cite{FF}, and it is based on the use of the first canonical conservation law and the first master symmetry. In the present paper, we apply the same 
ideas to the case of the lattice Boussinesq and lattice modified Boussinesq equations. 

By the lattice Boussinesq equation\footnote{An alternative version of a lattice Boussinesq equation was given in \cite{DJM} which could be viewed as a dimensional 
reduction of the modified lattice KP equation. However, it seems that this equation corresponds to the special case $p = \chi q$, where $\chi$ is a primitive root of 
unity, of equation (\ref{eq:dbsq}).}, we mean the following partial difference equation on the two-dimensional lattice
\begin{eqnarray}
& &\frac{\alpha^3-\beta^3}{\alpha-\beta+u_{n+1,m+1}-u_{n+2,m}}\,-\,
\frac{\alpha^3-\beta^3}{\alpha-\beta+u_{n,m+2}-u_{n+1,m+1}}\,-\,u_{n,m+1}u_{n+1,m+2} +\ u_{n+1,m}u_{n+2,m+1} \nn \\
&&+u_{n+2,m+2}\,(\alpha-\beta+u_{n+1,m+2}-u_{n+2,m+1})\ +\
u_{n,m}\,(\alpha-\beta+u_{n,m+1}-u_{n+1,m}) \nn \\
&=& (2\alpha+\beta)\,(u_{n+1,m}+u_{n+1,m+2})\ -\ (\alpha+2\beta)\,
(u_{n,m+1}+u_{n+2,m+1})\   \ ,  \label{eq:dbsq}
\end{eqnarray}
and we refer to
\begin{eqnarray}\label{eq:dmbsq} 
&& \left(\frac{\alpha^2 u_{n+1,m+1} - \beta^2 u_{n,m+2}}{\alpha u_{n,m+2} - \beta u_{n+1,m+1}} \right) \frac{u_{n+1,m+2}}{u_{n,m+1}} - \left( \frac{\alpha^2 u_{n+2,m} - \beta^2 u_{n+1,m+1}}{\alpha u_{n+1,m+1} - \beta u_{n+2,m}} \right) \frac{u_{n+2,m+1}}{u_{n+1,m}}  \nn \\
\nn \\
&& = \alpha\,\left(\frac{u_{n,m}}{u_{n+1,m}} - \frac{u_{n+1,m+2}}{u_{n+2,m+2}}\right) - \beta\,\left(\frac{u_{n,m}}{u_{n,m+1}} - \frac{u_{n+2,m+1}}{u_{n+2,m+2}}\right)
\end{eqnarray}
as the lattice modified Boussinesq equation. These equations were derived in \cite{NPCQ} in the context of the discrete Gel'fand-Dikii hierarchy, and, under appropriate continuum limits, they reduce to the continuous potential Boussinesq equation and the potential modified Boussinesq equation, respectively. These 9-point scalar 
equations can also be given as quadrilateral systems for three or two fields, respectively, and these forms will be used  in the subsequent analysis. 

In particular, our analysis will be focused on the derivation of hierarchies of symmetries and canonical conservation laws for these systems. These hierarchies are derived in a systematic way starting with the first generalized symmetry and the first canonical conservation law as seeds and making successive use of an extended generalized symmetry playing the role of a master symmetry. This procedure was used in \cite{X} in the derivation of corresponding hierarchies for all the ABS equations. 

Moreover, for the system related to the lattice modified Boussinesq equation, a corresponding system of partial differential equations (PDEs) is derived 
similarly to the case of the ABS equations \cite{TX}. These systems are also referred to as generating PDEs and were introduced in \cite{NJH} in connection 
with the lattice potential KdV equation. Subsequently, similar systems derived for the lattice Boussinesq equation \cite{TN2} and for all the ABS equations 
\cite{TX}. These systems inherit some properties of their lattice counterparts, more precisely they admit an auto-B{\"a}cklund transformation and a 
Lax pair, both of which actually follow from their multidimensional consistency. For the multidimensionally consistent generating PDE related to the lattice 
modified Boussinesq equation, we also present an auto-B{\"a}cklund transformation and a Lax pair.

The paper is organized as follows. The next section deals with the symmetries and conservation laws of the lattice Boussinesq system, while Section 
\ref{sec-mBsq} contains the corresponding results for the modified Boussinesq system. In Section \ref{sec-GPDE-mBsq} the generating PDE related to modified 
Boussinesq system is presented along with an auto-B{\"a}cklund transformation as well as a Lax pair for this system. The concluding section discusses some 
related points of current and future research.

\section{Lattice Boussinesq equation} \label{sec-Bsq}

In this section we study the generalized symmetries and conservation laws of the lattice Boussinesq equation. This equation was derived in \cite{NPCQ}, studied in \cite{W} while the corresponding generating PDE was presented in \cite{TN2}. Here we derive formulas for the recursive construction of symmetries and conservation laws using as seeds a pair of symmetries and a canonical conservation law. The interpretation of these symmetries as a hierarchy of differential-difference equations and our considerations lead naturally to a hierarchy of conservation laws and Lax pairs for these hierarchies as well.

Let us begin by writing the lattice Boussinesq equation as a system on a quadrilateral. Specifically, this equation can be written as a system involving the values of three fields ${\bf{u}}_{n,m} := (u_{n,m},v_{n,m},w_{n,m})$ on an elementary quadrilateral of the lattice which has the following form.
\begin{subequations} \label{bsq-sys-1}
\begin{eqnarray}
&& w_{n+1,m} - u_{n,m} u_{n+1,m}  + v_{n,m} = 0 \,,\\
&& w_{n,m+1} - u_{n,m} u_{n,m+1}  + v_{n,m} = 0 \,,\\
&& u_{n,m} u_{n+1,m+1} - v_{n+1,m+1} - w_{n,m}  - \frac{\alpha-\beta}{u_{n+1,m}-u_{n,m+1}}=0\,.
\end{eqnarray}
\end{subequations}
A Lax pair, gauge equivalent to the one derived in \cite{NPCQ}, cf. \cite{W, TN2}, is given by the following linear system for $\Psi_{n,m}$.
\begin{subequations} \label{dis-lp-bsq}
\begin{equation}
\Psi_{n+1,m}\,=\,L({\bf{u}}_{n,m},{\bf{u}}_{n+1,m};\alpha)\,\Psi_{n,m}\,,\quad
\Psi_{n,m+1}\,=\,L({\bf{u}}_{n,m},{\bf{u}}_{n,m+1};\beta)\,\Psi_{n,m}\,,
\end{equation}
where
\begin{equation} \label{bsq-L}
L({\bf{u}}_{n,m},{\bf{u}}_{n+1,m};\alpha)\,:=\,\frac{1}{(\alpha-\lambda)^{1/3}}\left(\begin{array}{ccc} -u_{n+1,m} & 1 & 0 \\ - v_{n+1,m} &0 &1 \\ \alpha-\lambda - u_{n,m} v_{n+1,m} + u_{n+1,m} w_{n,m} & -w_{n,m} & u_{n,m} \end{array} \right).
\end{equation}
\end{subequations}
The compatibility condition of the above linear system holds if and only if system (\ref{bsq-sys-1}) holds.

One useful property of system (\ref{bsq-sys-1}) is its covariance, i.e. its invariance under the mutual interchanges of shifted values of $u$, $v$ and $w$ accompanied with corresponding interchange of the parameters $\alpha$ and $\beta$. This property, which is obvious for the Lax pair (\ref{dis-lp-bsq}), allows us to present symmetries and conservation laws only for one lattice direction, specifically for the $n$ direction, while the results for the $m$ direction can follow by employing the interchanges
$$(u_{n+i,m+j},v_{n+i,m+j},w_{n+i,m+j},n,m,\alpha,\beta) \rightleftharpoons (u_{n+j,m+i},v_{n+j,m+i},w_{n+j,m+i},m,n,\beta,\alpha).$$

\begin{The}
The lattice Boussinesq system (\ref{bsq-sys-1}) admits a ten-dimensional algebra of Lie point symmetries generated by the vector fields
\begin{equation} \label{bsq-point-sym}
\begin{array}{l}
S_0 = u_{n,m} \partial_{u_{,m}} + 2 v_{n,m} \partial_{v_{n,m}} + 2 w_{n,m} \partial_{w_{n,m}} + 3 \alpha \partial_\alpha + 3 \beta \partial_\beta \,,\\
U_0 =  \partial_{u_{n,m}} +u_{n,m} \partial_{v_{n,m}} + u_{n,m} \partial_{w_{n,m}}\,,\\
T_0 =  \partial_{v_{n,m}} -  \partial_{w_{n,m}}\,,\\
S_k = \chi^{k (n+m)} u_{n,m} \partial_{u_{n,m}}- \chi^{k (n+m+2)} v_{n,m} \partial_{v_{n,m}} - \chi^{k (n+m+1)} w_{n,m} \partial_{w_{n,m}}\,,\\
U_k = \chi^{k(n+m)} \partial_{u_{n,m}} + \chi^{k (n+m+1)}u_{n,m} \partial_{v_{n,m}} + \chi^{k (n+m+2)} u_{n,m} \partial_{w_{n,m}}\,,\\
T_k = \chi^{k (n+m)}\partial_{v_{n,m}} - \chi^{k (n+m+2)} \partial_{w_{n,m}}\,,\\
E = \partial_\alpha + \partial_\beta\,,
\end{array}
\end{equation}
where $k=1,2$ and $\chi$ is a primitive root of unity $\chi^2 + \chi + 1 =0$.

Moreover, it admits a hierarchy of generalized symmetries $V_i$ and an extended generalized symmetry $G_1$ in the $n$ direction, which have the following form
\begin{equation} \label{bsq-gen-sym}
{V}_{i} ={\cal{R}}^{\,i-1} \left(\frac{1}{{r}_{n,m}}\right) \partial_{u_{n,m}} + {\cal{R}}^{\,i-1} \left(\frac{u_{n+1,m}}{{r}_{n,m}}\right) \partial_{v_{n,m}}  + {\cal{R}}^{\,i-1} \left(\frac{u_{n-1,m}}{{r}_{n,m}}\right) \partial_{w_{n,m}}\,,\quad i=1,2,\cdots,
\end{equation}
and 
\begin{equation} \label{bsq-m-sym}
 {G}_1 = \frac{n}{{r}_{n,m}} \partial_{u_{n,m}} + \frac{n\, u_{n+1,m}}{{r}_{n,m}} \partial_{v_{n,m}} + \frac{n\, u_{n-1,m}}{{r}_{n,m}} \partial_{w_{n,m}} + \partial_\alpha \,.
\end{equation}
Function $r_{n,m}$ and operator $\cal{R}$ are given by
\begin{subequations} \label{bsq-r-R}
\begin{equation}
r_{n,m} \,:=\, u_{n+1,m} u_{n-1,m} - v_{n+1,m} - w_{n-1,m}
\end{equation}
and
\begin{equation}
{\cal{R}} := \sum_{j=-\infty}^{\infty} \frac{j}{{r}_{n+j,m}} \partial_{u_{n+j,m}} + \sum_{j=-\infty}^{\infty} \frac{j\,u_{n+1+j,m}}{{r}_{n+j,m}} \partial_{v_{n+j,m}} +\sum_{j=-\infty}^{\infty} \frac{j\,u_{n-1+j,m}}{{r}_{n+j,m}} \partial_{w_{n+j,m}} + \partial_\alpha\,,
\end{equation}
\end{subequations}
respectively.

Moreover, system (\ref{bsq-sys-1}) admits a hierarchy of canonical conservation laws $\Delta_m \rho_{n,m}^{\,(k)} = \Delta_n \sigma_{n,m}^{\,(k)}$, the densities $\rho_{n,m}^{\,(k)}$ and the fluxes  $\sigma_{n,m}^{\,(k)}$ of which are defined recursively as
\begin{subeqnarray} \label{bsq-cl}
&& {\rho}_{n,m}^{\,(k)} = {{\cal{R}}}_\ast^{\,k}\left(\log \left( u_{n+1,m} u_{n-1,m} - v_{n+1,m} - w_{n-1,m}\right)^{\,2}\right)\,,\\ 
&& {\sigma}_{n,m}^{\,(k)} =  {{\cal{R}}}_\ast^{\,k} \left(\log \left(u_{n,m+1} u_{n-1,m} - v_{n,m+1} - w_{n-1,m} \right)^2\right)\,,
\end{subeqnarray}
for $k=0,1,\cdots$. Operator ${\cal{R}}_\ast$ is the ``negative part'' of $\cal{R}$, i.e.
\begin{equation} \label{bsq-R-}
{\cal{R}}_\ast := \sum_{j=-\infty}^{-1} \frac{j}{{r}_{n+j,m}} \partial_{u_{n+j,m}} + \sum_{j=-\infty}^{-1} \frac{j\,u_{n+1+j,m}}{{r}_{n+j,m}} \partial_{v_{n+j,m}} +\sum_{j=-\infty}^{-1} \frac{j\,u_{n-1+j,m}}{{r}_{n+j,m}} \partial_{w_{n+j,m}} + \partial_\alpha\,.
\end{equation}
\end{The}

\noindent {\bf{Proof}} It can be verified directly that the vector fields given in (\ref{bsq-point-sym}) are symmetry generators for system (\ref{bsq-sys-1}). They form a ten dimensional solvable Lie algebra with the non-vanishing commutators given by
\begin{eqnarray*}
&& [S_0,U_\ell] \,=\, -\, U_\ell\,,\quad [S_k,U_\ell] \,=\,-\, U_{k+\ell}\,\,{\rm{mod}}\,\,3 \,,\\
&& [S_0,T_\ell] \,=\, -2\,T_\ell\,,\quad [S_k,T_\ell] \,=\, \chi^{2 k} \,T_{k+\ell}\,\, {\rm{mod}}\,\,3\,,\\
&& [U_k,U_\ell] = (\chi^{\ell} - \chi^{k}) T_{k+\ell} \,\,{\rm{mod}}\,\,3\,,\\
&& [S_0,E] = -3 E\,,
\end{eqnarray*}
where $k =1,2$ and $\ell = 0,1,2$.

For the hierarchy (\ref{bsq-gen-sym}), we observe that ${G}_1$ is a master symmetry of ${V}_1$. Indeed, their commutator is non-zero, resulting to ${V}_2$, and $[V_1,[G_1,V_1]]=0$. This implies that we can construct recursively the whole hierarchy (\ref{bsq-gen-sym}) of commuting symmetries by setting $V_{i+1} = [G_1,V_i]$. Working out these commutators, relations (\ref{bsq-gen-sym}) easily follow.

Having constructed the hierarchy of generalized symmetries, one may derive hierarchies of canonical conservation laws starting with the first canonical conservation law $\Delta_m {\rho}_{n,m}^{\,(0)}=\Delta_n {\sigma}_{n,m}^{\,(0)}$. The density ${\rho}_{n,m}^{\,(0)}$ and the flux ${\sigma}_{n,m}^{\,(0)}$ have the form
\begin{subequations} \label{bsq-1-ccl}
\begin{equation}
{\rho}_{n,m}^{\,(0)} = \log {r}_{n,m}^{\,2}\quad {\mbox{and}} \quad {\sigma}_{n,m}^{\,(0)} = \log \left(u_{n,m+1} u_{n-1,m} - v_{n,m+1} - w_{n-1,m} \right)^2\,,
\end{equation}
respectively. The form of the conservation law suggests the introduction of the potential ${\phi}^{\,(0)}$ through the relations
\begin{equation}
\Delta_n {\phi}^{\,(0)}\,=\,{\rho}_{n,m}^{\,(0)}\,,\quad \Delta_n {\phi}^{\,(0)}\,=\,{\sigma}_{n,m}^{\,(0)} \,.
\end{equation}
\end{subequations}
If we consider the system constituted by (\ref{bsq-sys-1}) and (\ref{bsq-1-ccl}), then its lower order generalized symmetries follow by extending the action of ${V}_1$ and ${G}_1$ in the direction of the potential ${\phi}^{\,(0)}$. This means that they have the form
$${V}_1 = \frac{1}{{r}_{n,m}} \partial_{u_{n,m}}  + \frac{u_{n+1,m}}{{r}_{n,m}} \partial_{v_{n,m}} + \frac{u_{n-1,m}}{{r}_{n,m}} \partial_{w_{n,m}}  + {\Phi}_1^{(0)} \partial_{{\phi}^{\,(0)}}\,,$$
and
$${G}_1 = \frac{n}{{r}_{n,m}} \partial_{u_{n,m}} + \frac{n\,u_{n+1,m}}{{r}_{n,m}} \partial_{v_{n,m}} + \frac{n\,u_{n-1,m}}{{r}_{n,m}} \partial_{w_{n,m}}+ (n {\Phi}_1^{(0)}+{\phi}^{\,(1)}) \partial_{{\phi}^{\,(0)}} +\partial_\alpha\,,$$
respectively. The condition that these are symmetry generators for system (\ref{bsq-1-ccl}) implies that
$${\Phi}^{(0)}_1\,=\, {V}_1 \left( {\rho}^{\,(0)}_{n-1,m} \right)\,\equiv\,\left(\frac{1}{{r}_{n,m}} \partial_{u_{n,m}} + \frac{u_{n+1,m}}{{r}_{n,m}} \partial_{v_{n,m}}+ \frac{u_{n-1,m}}{{r}_{n,m}} \partial_{w_{n,m}}\right) {\rho}^{\,(0)}_{n-1,m}$$
and ${\phi}^{\,(1)}$ is the potential corresponding to the next conservation law. In particular, the density ${\rho}_{n,m}^{\,(1)}$ and the flux ${\sigma}_{n,m}^{\,(1)}$ of this conservation law are given by
\begin{equation}
{\rho}_{n,m}^{\,(1)} = {{\cal{R}}} \left({\rho}_{n,m}^{\,(0)}\right) - {\cal{S}}_n\left( {\Phi}^{(0)}_1 \right)\,,\quad 
{\sigma}_{n,m}^{\,(1)} =  {{\cal{R}}}_\ast \left({\sigma}_{n,m}^{\,(0)}\right)
\end{equation}
where ${{\cal{R}}}_\ast$ is the ``negative part'' of ${\cal{R}}$ given in (\ref{bsq-R-}).
 Moreover, taking into account the particular form of ${\Phi}^{(0)}_1$, density ${\rho}_{n,m}^{\,(1)}$ may also be written as ${\rho}_{n,m}^{\,(1)} = {{\cal{R}}}_\ast \left({\rho}_{n,m}^{\,(0)}\right) $.

If we include potential ${\phi}^{\,(1)}$ in our considerations and extend symmetries $V_1$ and $G_1$ to this direction, we will introduce a potential ${\phi}^{\,(2)}$ which corresponds to the next canonical conservation law. We can follow the same procedure with this new potential which will result to the introduction of another potential and corresponding conservation law. This procedure can be repeated infinitely many times, leading to a hierarchy of conservation laws. In particular, symmetry generators will have the form
\begin{eqnarray*}
&& {V}_1 = \frac{1}{{r}_{n,m}} \partial_{u_{n,m}} + \frac{u_{n+1,m}}{{r}_{n,m}} \partial_{v_{n,m}} + \frac{u_{n-1,m}}{{r}_{n,m}} \partial_{w_{n,m}}  + \sum_{k} {\Phi}_1^{(k)} \partial_{{\phi}^{\,(k)}}\,,\\
&& {G}_1 = \frac{n}{{r}_{n,m}} \partial_{u_{n,m}} + \frac{n\,u_{n+1,m}}{{r}_{n,m}} \partial_{v_{n,m}} + \frac{n\,u_{n-1,m}}{{r}_{n,m}} \partial_{w_{n,m}}+ \sum_k (n {\Phi}_1^{(k)}+{\phi}^{\,(k+1)}) \partial_{{\phi}^{\,(k)}} +\partial_\alpha\,,
\end{eqnarray*}
where
\begin{equation} \label{bsq-F}
{\Phi}^{(k)}_1 = {V}_1 \left( {\rho}^{\,(k)}_{n-1,m}\right) \equiv \left(\frac{1}{{r}_{n,m}} \partial_{u_{n,m}}+ \frac{u_{n+1,m}}{{r}_{n,m}} \partial_{v_{n,m}} + \frac{u_{n-1,m}}{{r}_{n,m}} \partial_{w_{n,m}}\right) {\rho}^{\,(k)}_{n-1,m} .
\end{equation}
Using the last relation, the corresponding densities and fluxes of the conservation laws $\Delta_m {\rho}^{\,(k)}_{n,m} =  \Delta_n {\sigma}^{\,(k)}_{n,m}$  can be written as in (\ref{bsq-cl}).

Moreover, it is straightforward to derive the action of ${V}_i$, $i >1$, on the potentials ${\phi}^{(k)}$ using the definition of these symmetries and the above extended forms of ${V}_1$ and ${G}_1$.  This derivation results to
\begin{subequations}
\begin{equation}
{V}_{i} ={\cal{R}}^{\,i-1} \left(\frac{1}{{r}_{n,m}}\right) \partial_{u_{n,m}} + {\cal{R}}^{\,i-1} \left(\frac{u_{n+1,m}}{{r}_{n,m}}\right) \partial_{v_{n,m}}  + {\cal{R}}^{\,i-1} \left(\frac{u_{n-1,m}}{{r}_{n,m}}\right) \partial_{w_{n,m}} + \sum_k {\Phi}_i^{(k)} \partial_{{\phi}^{(k)}},
\end{equation}
which is valid for $i=1,2,\cdots$, and
$${\Phi}^{(k)}_i = \sum_{j=0}^{i-1} \left(\begin{array}{c} i-1 \\ j \end{array} \right) (-1)^{i+j-1} {\cal{R}}^{\,j} \left({\Phi}_1^{(i+k-j-1)}\right).$$
Using relations (\ref{bsq-F}) and (\ref{bsq-cl}), the above relation can be written also as
\begin{equation} \label{bsq-Fk}
{\Phi}^{(k)}_i = \sum_{j=0}^{i-1} \left(\begin{array}{c} i-1 \\ j \end{array} \right) (-1)^{i+j-1} {\cal{R}}^{\,j} \circ {V}_1 \circ {\cal{S}}_n^{-1} \circ {{\cal{R}}}_\ast^{\,k+i-j-1} \left( \log {r}_{n,m}^{\,2}\right),
\end{equation}
\end{subequations}
which hold for $i=1,2,\cdots$ and $k=0,1,2,\cdots$.
\hfill $\Box$ \\

The interpretation of symmetries as differential-difference equations commuting with the lattice system and the previous analysis lead to hierarchies of differential-difference equations and corresponding conservation laws. Since they involve shifts only in one lattice direction, in what follows we drop the second index $m$ which does not vary. In this setting, we have the following 
\begin{Cor}
The hierarchy of differential-difference equations
\begin{equation} \label{dd-bsq-sys}
\frac{\partial u_n}{\partial t_i}={\cal{R}}^{\,i-1}\left( \frac{1}{{r}_n} \right),\quad
\frac{\partial v_n}{\partial t_i} = {\cal{R}}^{\,i-1}\left( \frac{u_{n+1}}{{r}_n} \right),\quad 
\frac{\partial w_n}{\partial t_i} = {\cal{R}}^{\,i-1}\left( \frac{u_{n-1}}{{r}_n} \right), \quad i =1,2,\cdots,
\end{equation}
where 
\begin{subequations}
\begin{equation}
r_n := u_{n+1} u_{n-1} - v_{n+1} - w_{n-1}
\end{equation}
and
\begin{equation} \label{dd-bsq-R}
{\cal{R}} := \sum_{j=-\infty}^{\infty} \frac{j}{{r}_{n+j}} \partial_{u_{n+j}} + \sum_{j=-\infty}^{\infty} \frac{j\,u_{n+1+j}}{{r}_{n+j}} \partial_{v_{n+j}} +\sum_{j=-\infty}^{\infty} \frac{j\,u_{n-1+j}}{{r}_{n+j}} \partial_{w_{n+j}} + \partial_\alpha\,,
\end{equation}
\end{subequations}
admits a master symmetry 
\begin{equation}  \label{dd-bsq-m-sym}
\frac{\partial u_n}{\partial \alpha} \,=\,\frac{n}{{r}_{n}} ,\quad \frac{\partial v_n}{\partial \alpha} \,=\,\frac{n\,u_{n+1}}{{r}_{n}},\quad
\frac{\partial w_n}{\partial \alpha} \,=\,\frac{n\,u_{n-1}}{{r}_{n}}.
\end{equation}

Moreover, systems (\ref{dd-bsq-sys}) admit a hierarchy of canonical conservation laws, which have the following form
$$\frac{\partial \phantom{t_i}}{\partial t_i}\left({{\cal{R}}}_\ast^{\,k}\left(\log {r}_{n}^{\,2}\right)\right)\,=\,\Delta_n \left(\sum_{j=0}^{i-1} \left(\begin{array}{c} i-1 \\ j \end{array} \right) (-1)^{i+j-1} {\cal{R}}^{\,j} \circ {\cal{V}} \circ {\cal{S}}_n^{-1} \circ {{\cal{R}}}_\ast^{\,k+i-j-1} \left( \log {r}_{n}^{\,2}\right) \right),$$
with $k=0,1,2,\cdots$, $i=1,2,\cdots$. In the above relation, ${\cal{R}}_\ast$ is the ``negative part'' of $\cal{R}$ given in (\ref{dd-bsq-R}) and 
$${\cal{V}} := \frac{1}{r_n} \partial_{u_n} + \frac{u_{n+1}}{r_n} \partial_{v_n} + \frac{u_{n-1}}{r_n} \partial_{w_n}.$$

Finally, a Lax pair for systems (\ref{dd-bsq-sys}) and (\ref{dd-bsq-m-sym}) is given by
\begin{subequations} \label{dd-lp-bsq}
\begin{eqnarray}
&& \Psi_{n+1}\,=\,L({\bf{u}}_{n},{\bf{u}}_{n+1};\alpha)\,\Psi_{n}\,, \\
&& \frac{\partial \Psi_{n}}{\partial t_i}\,=\,\left ({\cal{R}}^{\,i-1} M({\bf{u}}_{n},{\bf{u}}_{n+1},{\bf{u}}_{n-1};\alpha)\right)\,\Psi_{n}\,,\quad i=1,2,\cdots\,,\\
&& \frac{\partial \Psi_{n}}{\partial \alpha}\,=\,n\, M({\bf{u}}_{n},{\bf{u}}_{n+1},{\bf{u}}_{n-1};\alpha)\,\Psi_{n}\,. \label{dd-lp-bsq-ms}
\end{eqnarray}
Matrix $L$ is defined in (\ref{bsq-L}) and traceless matrix $M$ \footnote{The continuous part (\ref{dd-lp-bsq-ms}) of the Lax pair for the master symmetry (\ref{dd-bsq-m-sym}) is gauge equivalent to the one derived in \cite{TN2}.} is given by
\begin{eqnarray} \label{dd-bsq-M}
&& M({\bf{u}}_{n},{\bf{u}}_{n+1},{\bf{u}}_{n-1};\alpha)\,:=\nonumber\\ 
&& \qquad \frac{1}{\alpha-\lambda}\,\frac{1}{{r}_n}\left(\begin{array}{ccc} 
\frac{v_{n+1} -2 w_{n-1} - u_{n-1} u_{n+1}}{3} & u_{n-1} & -1 \\
-u_{n+1} w_{n-1} & \frac{v_{n+1} + w_{n-1} +2 u_{n-1} u_{n+1}}{3} & -u_{n+1} \\
-v_{n+1} w_{n-1} & u_{n-1} v_{n+1} & \frac{w_{n-1}-2 v_{n+1} - u_{n-1} u_{n+1}}{3}
 \end{array} \right)\,.
\end{eqnarray}
\end{subequations}
\end{Cor}

Here we have constructed infinite sequences of symmetries and conservation laws for system (\ref{bsq-sys-1}) which is related to the lattice Boussinesq equation (\ref{eq:dbsq}). In the next section we are going to establish similar results for the lattice modified Boussinesq equation (\ref{eq:dmbsq}) written as a system for two fields.

\section{Lattice modified Boussinesq equation} \label{sec-mBsq}

The modified Boussinesq equation was derived in \cite{NPCQ} and studied in \cite{W}, while its Lie point and lower order generalized symmetries were presented in \cite{TN1}. In this section we present hierarchies of symmetries and canonical conservation laws for this equation, as well as  corresponding structures for differential-difference equations. We follow the derivation of the previous section and, thus, we omit most of the details here and present only the necessary results.

The lattice modified Boussinesq equation can be written as a system for two fields ${\bf{u}}_{n,m} := (u_{n,m},v_{n,m})$ defined on an elementary quadrilateral of the lattice in the following form  
\begin{subequations} \label{mbsq-sys}
\begin{eqnarray}
u_{n+1,m+1} &=& v_{n,m}\,\frac{\alpha u_{n,m+1} - \beta u_{n+1,m}}{\alpha v_{n+1,m} - \beta v_{n,m+1}}\,, \\
v_{n+1,m+1} &=& \frac{v_{n,m}}{u_{n,m}}\,\frac{\alpha u_{n+1,m} v_{n,m+1} - \beta u_{n,m+1} v_{n+1,m}}{\alpha v_{n+1,m} - \beta v_{n,m+1}}\,, 
\end{eqnarray}
\end{subequations}
cf. \cite{N1} where a similar 2-field system for the lattice modified Boussinesq equation was first given. 
This system is the necessary and sufficient condition for the consistency of the following linear system, which, in this sense, constitutes a Lax pair for system (\ref{mbsq-sys}).
\begin{subequations}\label{lax-pair-dis-sys-mbsq}
\begin{equation}
\Psi_{n+1,m} = L({\bf{u}}_{n,m},{\bf{u}}_{n+1,m};\alpha) \Psi_{n,m} \,,\quad \Psi_{n,m+1} = L({\bf{u}}_{n,m},{\bf{u}}_{n,m+1};\beta) \Psi_{n,m}\,,
\end{equation}
where
\begin{equation} \label{mbsq-L}
L({\bf{u}}_{n,m},{\bf{u}}_{n+1,m};\alpha)  := 
\frac{1}{(\alpha^3-\lambda^3)^{1/3}} \left(\begin{array}{ccc} 
\frac{\alpha v_{n+1,m}}{v_{n,m}} & 0 & -\lambda\\
- \lambda & \frac{\alpha u_{n,m}}{u_{n+1,m}} & 0 \\
0 & - \lambda & \frac{\alpha u_{n+1,m} v_{n,m}}{u_{n,m} v_{n+1,m}}
\end{array} \right)\,.
\end{equation}
\end{subequations}

The point symmetries of this system, which were given in \cite{TN1},
\begin{subequations}
\begin{eqnarray}
&& T_u = u_{n,m} \partial_{u_{n,m}}\,,\quad T_v = v_{n,m} \partial_{v_{n,m}}\,, \nonumber \\
&& S(\chi) = \chi^{n+m} u_{n,m} \partial_{u_{n,m}} - \chi^{n+m+1} v_{n,m} \partial_{v_{n,m}}\,,
\end{eqnarray}
where $\chi$ is a primitive root of unit, $\chi^2+\chi+1=0$. Additionally, system (\ref{mbsq-sys}) admits an extended point symmetry, namely
\begin{equation}
E = \alpha \partial_\alpha + \beta \partial_\beta\,.
\end{equation}
\end{subequations}

\begin{The}
The modified Boussinesq system (\ref{mbsq-sys}) admits a hierarchy of generalized symmetries $V_i$ and an extended generalized symmetry $G_1$ which have the form
\begin{subequations} \label{mbsq-sym}
\begin{equation} \label{mbsq-V}
{V}_i\,=\,{\cal{P}}^{\,i-1}\left({R}_1\right)\partial_{u_{n,m}} + {\cal{P}}^{\,i-1}\left({P}_1\right)\partial_{v_{n,m}}\,,\quad i=1,2,\cdots,
\end{equation}
and
\begin{equation} \label{mbsq-G}
{G}_1 = n\,{R}_1 \partial_{u_{n,m}} + n\,{P}_1 \partial_{v_{n,m}}\,-\,\alpha\,\partial_\alpha\,,
\end{equation}
respectively. In the above relations,
\begin{eqnarray}
&& {R}_1:=\frac{3 u_{n,m} u_{n+1,m} v_{n,m}}{u_{n+1,m} v_{n,m} + u_{n,m} v_{n-1,m} + u_{n-1,m} v_{n+1,m}}-u_{n,m}\,,\\
&&{{P}}_1 := \frac{-3 u_{n,m} v_{n-1,m} v_{n,m}}{u_{n+1,m} v_{n,m} + u_{n,m} v_{n-1,m} + u_{n-1,m} v_{n+1,m}}+v_{n,m}\,,
\end{eqnarray}
and
\begin{equation}
{\cal{P}}\,:=\,\sum_{j=-\infty}^{\infty} j\, {\cal{S}}_n^j({R}_1) \partial_{u_{n+j,m}} + \sum_{j=-\infty}^{\infty} j\,{\cal{S}}_n^j({P}_1) \partial_{v_{n+j,m}} -\alpha \partial_\alpha\,.
\end{equation}
\end{subequations}

Moreover, it admits a hierarchy of canonical conservation laws $\Delta_m \rho_{n,m}^{\,(k)} = \Delta_n \sigma_{n,m}^{\,(k)}$, the densities $\rho_{n,m}^{\,(k)}$ and the fluxes  $\sigma_{n,m}^{\,(k)}$ of which are given by
\begin{subeqnarray} \label{mbsq-cl}
&& {\rho}_{n,m}^{(k)}\,=\,\Big( {{\cal{P}}}_\ast + {\cal{D}}\Big)^k \left(\log\left(\frac{u_{n+1,m} v_{n,m} + u_{n,m} v_{n-1,m} + u_{n-1,m} v_{n+1,m}}{u_{n,m} v_{n,m}}\right)^2\right)\,,\\
&& {\sigma}^{(k)}_{n,m}\,=\,{{\cal{P}}}_\ast^k \left(\log\left(\frac{\alpha^2 u_{n,m} v_{n-1,m} + \alpha \beta u_{n,m} v_{n,m+1} + \beta^2 u_{n,m+1} v_{n,m}}{(\alpha^3-\beta^3) u_{n,m} v_{n,m}}\right)^2\right)\,,
\end{subeqnarray}
for $k=0,1,2,\cdots$. In the above relations, operators $\cal{D}$ and ${\cal{P}}_\ast$ are given by
$${\cal{D}} := 2 u_{n+1,m} \partial_{u_{n+1,m}} +v_{n+1,m} \partial_{v_{n+1,m}} $$
 and 
\begin{equation}
{\cal{P}}_\ast\,:=\,\sum_{j=-\infty}^{-1} j\, {\cal{S}}_n^j({R}_1) \partial_{u_{n+j,m}} + \sum_{j=-\infty}^{-1} j\,{\cal{S}}_n^j({P}_1) \partial_{v_{n+j,m}} -\alpha \partial_\alpha\,,
\end{equation}
respectively.
\end{The}

\noindent {\bf{Proof}} Using the fact that $G_1$, given in (\ref{mbsq-G}), is a master symmetry for $V_1$, hierarchy (\ref{mbsq-V})  can be constructed recursively by setting $V_{i+1} = [G_1,V_i]$ for $i=1,2,\cdots$.

The hierarchy of conservation laws with densities and fluxes given in (\ref{mbsq-cl}) can be constructed using the same method and analysis we used in the previous section. That is starting with
\begin{eqnarray*}
&& {\rho}_{n,m}^{(0)}\,=\,\log\left(\frac{u_{n+1,m} v_{n,m} + u_{n,m} v_{n-1,m} + u_{n-1,m} v_{n+1,m}}{u_{n,m} v_{n,m}}\right)^2,\\
&& {\sigma}^{(0)}_{n,m}\,=\,\log\left(\frac{\alpha^2 u_{n,m} v_{n-1,m} + \alpha \beta u_{n,m} v_{n,m+1} + \beta^2 u_{n,m+1} v_{n,m}}{(\alpha^3-\beta^3) u_{n,m} v_{n,m}}\right)^2,
\end{eqnarray*}
and employing the first symmetry ${V}_1$ and its master symmetry $G_1$. In the course of the construction, we extend symmetries $V_1$ and $V_i$, $i > 1$, in the direction of the corresponding potentials and these extensions are given by the functions 
\begin{subequations}
\begin{equation}
{\Phi}^{(k)}_1 = \left(({{R}}_1-2 u_{n,m}) \partial_{u_{n,m}} + ({P}_1 - v_{n,m}) \partial_{v_{n,m}} \right) {\rho}^{\,(k)}_{n-1,m}\,,\quad k =0,1,\cdots , 
\end{equation}
and
\begin{equation}
{\Phi}^{(k)}_i = \sum_{j=0}^{i-1} \left(\begin{array}{c} i-1 \\ j \end{array} \right) (-1)^{i+j-1} {\cal{P}}^{\,j} \left({\Phi}_1^{(i+k-j-1)}\right),\quad  k =0,1,\cdots,
\end{equation}
\end{subequations}
respectively. \hfill $\Box$

\noindent {\bf{Remark}} It is worth mentioning that the first generalized symmetry ${V}_1$ is actually a linear combination of two point symmetries and a generalized one, specifically
$${V}_1 = -u_{n,m} \partial_{u_{n,m}} + v_{n,m} \partial_{v_{n,m}} + 3 \left(\frac{u_{n,m} u_{n+1,m} v_{n,m}}{{r}_{n,m}}\partial_{u_{n,m}} +  \frac{- u_{n,m} v_{n-1,m} v_{n,m}}{{r}_{n,m}}\partial_{v_{n,m}}\right)\,.$$
Our choice for the form of ${V}_1$ is motivated from the corresponding form of ${G}_1$ and does not affect our analysis.\\

The interpretation of symmetries as differential-difference equations leads to 

\begin{Cor}
The hierarchy of differential-difference equations
\begin{equation} \label{mbsq-dd}
\frac{\partial u_{n}}{\partial t_i}\,=\,{\cal{P}}^{\,i-1} ({R}_1)\,,\quad \frac{\partial v_{n}}{\partial t_i}\,=\,{\cal{P}}^{\,i-1}({P}_1)\,,\quad i=1,2,\cdots,
\end{equation}
admits a master symmetry
\begin{equation} \label{mbsq-dd-ms}
\alpha \frac{\partial u_{n}}{\partial \alpha}\,=\,-n\,{{R}}_1\,,\quad \alpha \frac{\partial v_{n}}{\partial \alpha}\,=\,-n\,{{P}}_1\,.
\end{equation}
In the above relations,
\begin{subequations}
\begin{eqnarray}
&& {R}_1:=\frac{3 u_{n} u_{n+1} v_{n}}{u_{n+1} v_{n} + u_{n} v_{n-1} + u_{n-1} v_{n+1}}-u_{n}\,,\\
&&{{P}}_1 := \frac{-3 u_{n} v_{n-1} v_{n}}{u_{n+1} v_{n} + u_n v_{n-1} + u_{n-1} v_{n+1}}+v_{n}\,,
\end{eqnarray}
and
\begin{equation} \label{mbsq-P}
{\cal{P}}\,:=\,\sum_{j=-\infty}^{\infty} j\, {\cal{S}}_n^j({R}_1) \partial_{u_{n+j}} + \sum_{j=-\infty}^{\infty} j\,{\cal{S}}_n^j({P}_1) \partial_{v_{n+j}} -\alpha \partial_\alpha\,.
\end{equation}
\end{subequations}

Moreover, it admits a hierarchy of canonical conservation laws $\partial_{t_i} \rho^{(k)}_n = \Delta_n \Phi_i^{(k)}$, the densities of which are given by 
$${\rho}_{n}^{(k)}\,=\,\Big( {{\cal{P}}}_\ast + {\cal{D}}\Big)^k \left(\log\left(\frac{u_{n+1} v_{n} + u_{n} v_{n-1} + u_{n-1} v_{n+1}}{u_{n} v_{n}}\right)^2\right)\,,\quad k = 0,1,2,\cdots,$$
and the corresponding fluxes are defined recursively by
\begin{eqnarray*}
&& {\Phi}^{(k)}_1 = \Big(({{R}}_1-2 u_{n}) \partial_{u_{n}} + ({P}_1 - v_{n}) \partial_{v_{n}} \Big) {\rho}^{\,(k)}_{n-1}\,, \\
&& {\Phi}^{(k)}_i = \sum_{j=0}^{i-1} \left(\begin{array}{c} i-1 \\ j \end{array} \right) (-1)^{i+j-1} {\cal{P}}^{\,j} \left({\Phi}_1^{(i+k-j-1)}\right),\quad i >1,
\end{eqnarray*}
for $k=0,1,2,\cdots$. In the above relations, 
$${\cal{D}} := 2 u_{n+1} \partial_{u_{n+1}} +v_{n+1} \partial_{v_{n+1}} $$
and ${\cal{P}}_\ast$ is the negative part of $\cal{P}$ defined in (\ref{mbsq-P}), i.e.
$$
{\cal{P}}_\ast\,:=\,\sum_{j=-\infty}^{-1} j\, {\cal{S}}_n^j({R}_1) \partial_{u_{n+j}} + \sum_{j=-\infty}^{-1} j\,{\cal{S}}_n^j({P}_1) \partial_{v_{n+j}} -\alpha \partial_\alpha\,.
 $$

A Lax pair for the differential-difference hierarchy (\ref{mbsq-dd}) and its master symmetry (\ref{mbsq-dd-ms}) is given by
\begin{subeqnarray}\label{lax-pair-dd-mbsq}
&& \Psi_{n+1} = L({\bf{u}}_{n},{\bf{u}}_{n+1};\alpha) \Psi_{n} \,,\\
&& \frac{\partial \Psi_n}{\partial t_k} = \left({\cal{P}}^{k-1}_n M({\bf{u}}_n,{\bf{u}}_{n+1},{\bf{u}}_{n-1};\alpha)\right) \Psi_n\,,\\
&& \frac{\partial \Psi_n}{\partial \alpha} = -\frac{n}{\alpha} \left(M({\bf{u}}_n,{\bf{u}}_{n+1},{\bf{u}}_{n-1};\alpha)  + \frac{\alpha^3}{\alpha^3-\lambda^3} {\rm{I}} \right) \Psi_n\,,
\end{subeqnarray}
where matrix $L$ is defined in (\ref{mbsq-L}) and
\begin{equation}
M({\bf{u}}_n,{\bf{u}}_{n+1},{\bf{u}}_{n-1};\alpha) := 
\frac{-3 \alpha^3}{(\alpha^3-\lambda^3) {r}_{n}} \left(\begin{array}{ccc} 
u_{n} v_{n-1} & \frac{\lambda^2 u_{n} v_{n}}{\alpha^2} & \frac{\lambda u_{n-1} v_{n}}{\alpha}\\
\frac{\lambda u_{n+1} v_{n-1}}{\alpha} & u_{n+1} v_{n} & \frac{\lambda^2 u_{n-1} u_{n+1} v_{n}}{\alpha^2 u_{n}} \\
\frac{\lambda^2 u_{n} v_{n-1} v_{n+1}}{\alpha^2 v_{n}} & \frac{\lambda u_{n} v_{n+1}}{\alpha} & u_{n-1} v_{n+1}
\end{array} \right).
\end{equation}
\end{Cor}

Thus we have systematically constructed the generalized symmetries and the canonical conservation laws for both systems (\ref{bsq-sys-1}) and (\ref{mbsq-sys}) which are related to lattice potential and modified Boussinesq equations, respectively. The existence of such hierarchies establishes the integrability of these systems, the multidimensional consistency of which was established in \cite{W}. In the next section we employ both of master symmetries of system (\ref{mbsq-sys}) in the derivation of an integrable and multidimensionally consistent system of PDEs.

\section{Continuous symmetric reduction of the lattice modified Boussinesq system} \label{sec-GPDE-mBsq}

The concept of generating PDE was introduced in \cite{NJH} and a systematic method for the derivation of such systems was presented in 
\cite{TX} based on the symmetries of the underlying discrete system. Here we apply this method to the case of the lattice modified 
Boussinesq system (\ref{mbsq-sys}) using both of its master symmetries. That is, we look for solutions of system (\ref{mbsq-sys}) which remain 
invariant under the action of both symmetries, and which lead to a system of PDEs rather than a system of ordinary difference equations
\footnote{We point out that in this respect such symmetry reductions are different from reductions by Lie point  
symmetries, in that the latter usually reduce the space of solutions to a finite-dimensional one, whereas with the reductions we impose here the 
solution space remains infinite-dimensional.}. 

More precisely, introducing the notation
\begin{eqnarray*}
&& u := u_{n,m},\quad u_1 := u_{n+1,m},\quad u_2 := u_{n,m+1},\quad u_{-1} := u_{n-1,m},\quad u_{-2} := u_{n,m-1}\,,\\
&&v := v_{n,m},\quad v_1 := v_{n+1,m},\quad v_2 := v_{n,m+1},\quad v_{-1} := v_{n-1,m},\quad v_{-2} := v_{n,m-1}\,,
\end{eqnarray*}
we impose that $u$, $v$ and their shifts satisfy, additionally to system (\ref{mbsq-sys}), the ``similarity constraints'' 
\begin{subequations}
\begin{eqnarray}
&& \alpha \frac{\partial u}{\partial \alpha} \,+\,n\,u\,\left(\frac{3\,v\,u_1}{u v_{-1} + u_1 v + u_{-1} v_1}-1 \right)\,=\,0\,,\\
&& \alpha  \frac{\partial v}{\partial \alpha} \,+\,n\,v\,\left(1\,-\,\frac{3\,u\,v_{-1}}{u v_{-1} + u_1 v + u_{-1} v_1} \right)\,=\,0\,,\\
&& \beta \frac{\partial u}{\partial \beta} \,+\,m\,u\,\left(\frac{3\,v\,u_2}{u v_{-2} + u_2 v + u_{-2} v_2}-1 \right)\,=\,0\,,\\
&& \beta \frac{\partial v}{\partial \beta} \,+\,m\,v\,\left(1\,-\,\frac{3\,u\,v_{-2}}{u v_{-2} + u_2 v + u_{-2} v_2} \right)\,=\,0\,, 
\end{eqnarray}
\end{subequations}
which in \cite{W} (in a different notation) were presented as a system of differential-difference equations compatible with the original 
lattice system (\ref{mbsq-sys}).   

From the above system of differential-difference equations accompanied with the lattice modified Boussinesq system (\ref{mbsq-sys}), a system of partial differential equations for $({\bf{u}},{\bf{u}}_1,{\bf{u}}_2) := (u,v,u_1,v_1,u_2,v_2)$ can be derived in a systematic way as it was done for the ABS equations in \cite{TX}. This system has the following form
\begin{subequations} \label{con-sys-mbsq}
\begin{eqnarray}
\frac{\partial u_1}{\partial \beta} &=& \frac{\alpha (\alpha u_2-\beta u_1) (\alpha u_1 v_2-\beta u_2 v_1)}{(\alpha^3 - \beta^3) u u_2 v_2}\frac{\partial u}{\partial \beta} - \frac{\beta (\alpha u_2-\beta u_1) (\beta v_2-\alpha v_1)}{(\alpha^3 - \beta^3) v v_2}\frac{\partial v}{\partial \beta}\nonumber  \\
&& - m\,\frac{\alpha (u_2 v_1 (\alpha u_2-\beta u_1) + v_2 (\beta u_2^2 - \alpha u_1^2))}{(\alpha^3-\beta^3) u_2 v_2} \,,\\
\frac{\partial v_1}{\partial \beta} &=& \frac{\beta (\beta v_2-\alpha v_1) (\alpha u_1 v_2-\beta u_2 v_1)}{(\alpha^3 - \beta^3) u u_2 v_2}\frac{\partial u}{\partial \beta} - \frac{(\alpha v_1-\beta v_2) (\beta^2 v_1-\alpha^2 v_2)}{(\alpha^3 - \beta^3) v v_2}\frac{\partial v}{\partial \beta}\nonumber  \\
&& - m\,\frac{\alpha (u_1 v_2 (\beta v_2-\alpha v_1) + u_2 (\alpha v_2^2 - \beta v_1^2))}{(\alpha^3-\beta^3) u_2 v_2}\,,\\
\frac{\partial u_2}{\partial \alpha} &=& \frac{\beta (\beta u_1-\alpha u_2) (\beta u_2 v_1-\alpha u_1 v_2)}{(\beta^3 - \alpha^3) u u_1 v_1}\frac{\partial u}{\partial \alpha} - \frac{\alpha (\beta u_1-\alpha u_2) (\alpha v_1-\beta v_2)}{(\beta^3 - \alpha^3) v v_1}\frac{\partial v}{\partial \alpha}\nonumber  \\
&& - n\,\frac{\beta (u_1 v_2 (\beta u_1-\alpha u_2) + v_1 (\alpha u_1^2 - \beta u_2^2))}{(\beta^3-\alpha^3) u_1 v_1} \,,\\
\frac{\partial v_2}{\partial \alpha} &=& \frac{\alpha (\alpha v_1-\beta v_2) (\beta u_2 v_1-\alpha u_1 v_2)}{(\beta^3 - \alpha^3) u u_1 v_1}\frac{\partial u}{\partial \alpha} - \frac{(\beta v_2-\alpha v_1) (\alpha^2 v_2-\beta^2 v_1)}{(\beta^3 - \alpha^3) v v_1}\frac{\partial v}{\partial \alpha}\nonumber  \\
&& - n\,\frac{\beta (u_2 v_1 (\alpha v_1-\beta v_2) + u_1 (\beta v_1^2 - \alpha v_2^2))}{(\beta^3-\alpha^3) u_1 v_1}\,,\\
\frac{\partial^2 u}{\partial \alpha \partial \beta} &=& \frac{1}{u} \left( 1 + \frac{\alpha \beta}{\alpha^3 - \beta^3}\left(\beta \frac{u_1 v_2}{u_2 v_1} - \alpha  \frac{u_2 v_1}{u_1 v_2}\right) \right) \frac{\partial u}{\partial \alpha} \frac{\partial u}{\partial \beta} \nonumber \\
&& +\frac{\alpha \beta}{\alpha^3 - \beta^3} \frac{\alpha v_1 - \beta v_2}{v} \left(\frac{u_2}{u_1 v_2} \frac{\partial u}{\partial \alpha} \frac{\partial v}{\partial \beta} + \frac{u_1}{u_2 v_1} \frac{\partial u}{\partial \beta} \frac{\partial v}{\partial \alpha} \right) \nonumber \\
&&  - m \frac{\alpha}{\alpha^3-\beta^3}\left(\left(\alpha \frac{u_2 v_1}{u_1 v_2} + \beta \frac{u_1v_2}{u_2v_1} + \beta \frac{u_2}{u_1} \right)\frac{\partial u}{\partial \alpha} - \frac{u u_1 (\beta v_2-\alpha v_1)}{u_2 v v_1} \frac{\partial v}{\partial \alpha}\right) \nonumber \\
&&  - n \frac{\beta}{\beta^3-\alpha^3}\left(\left(\beta  \frac{u_1v_2}{u_2v_1} + \alpha \frac{u_2 v_1}{u_1 v_2} + \alpha \frac{u_1}{u_2} \right) \frac{\partial u}{\partial \beta} - \frac{u u_2 (\alpha v_1-\beta v_2)}{u_1 v v_2} \frac{\partial v}{\partial \beta}\right) \nonumber \\
&& + \frac{n m}{\alpha^3 - \beta^3} \frac{(\beta v_2 + \alpha v_1) (u_2^2 v_1 - u_1^2 v_2)}{u_1 u_2 v_1 v_2}\,, \\
\frac{\partial^2 v}{\partial \alpha \partial \beta} &=& \frac{1}{\alpha^3 - \beta^3} \frac{(\alpha^2 v_2 + \beta^2 v_1) (\alpha v_1 - \beta v_2)}{v v_1 v_2} \frac{\partial v}{\partial \alpha} \frac{\partial v}{\partial \beta} \nonumber \\
&& +\frac{\alpha \beta}{\alpha^3 - \beta^3} \frac{\alpha u_1 v_2 - \beta u_2 v_1}{u} \left(\frac{1}{u_2 v_2} \frac{\partial v}{\partial \alpha} \frac{\partial u}{\partial \beta} + \frac{1}{u_1 v_1}  \frac{\partial v}{\partial \beta} \frac{\partial u}{\partial \alpha} \right) \nonumber \\
&&  - m \frac{\alpha}{\alpha^3-\beta^3}\left(\left( \alpha \frac{u_1}{u_2} + \alpha \frac{v_2}{v_1} + \beta \frac{v_1}{v_2} \right)\frac{\partial v}{\partial \alpha} - \frac{v (\alpha u_1 v_2-\beta u_2 v_1)}{u u_1 v_1} \frac{\partial u}{\partial \alpha}\right) \nonumber \\
&&  - n \frac{\beta}{\beta^3-\alpha^3}\left(\left( \beta \frac{u_2}{u_1} + \beta \frac{v_1}{v_2} + \alpha \frac{v_2}{v_1} \right)\frac{\partial v}{\partial \beta} - \frac{v (\beta u_2 v_1-\alpha u_1 v_2)}{u u_2 v_2} \frac{\partial u}{\partial \beta}\right) \nonumber \\
&& + \frac{n m}{\alpha^3 - \beta^3} \frac{v (\alpha u_1 v_2+\beta u_2 v_1) (u_2 v_2 - u_1 v_1)}{u_1 u_2 v_1 v_2} \,.
\end{eqnarray}
\end{subequations}
In the following we will denote this system by $B_{n,m}({\bf{u}},{\bf{u}}_1,{\bf{u}}_2;\alpha,\beta)$ or, simply, by $B_{n,m}$. 

The integrability properties of this system are inherited to it by its lattice counterpart and are the following ones : it is multidimensionally consistent in the sense of \cite{XNL}, it admits  an auto-B{\"a}cklund transformation and a Lax pair. These properties are formulated in the following three theorems which can be proven by straightforward calculations.

\begin{The}
The three copies
$$B_{n_i,n_j}({\bf{u}},{\bf{u}}_i,{\bf{u}}_j;\alpha_i,\alpha_j),\quad B_{n_j,n_k}({\bf{u}},{\bf{u}}_j,{\bf{u}}_k;\alpha_j,\alpha_k),\quad B_{n_k,n_i}({\bf{u}},{\bf{u}}_k,{\bf{u}}_i;\alpha_k,\alpha_i) $$
of the above system are consistent with each other. That is, the three different ways to evaluate $\partial_{\alpha_i} \partial_{\alpha_j} \partial_{\alpha_k} {\bf{u}}$ and the two different ways to evaluate $\partial_{\alpha_i} \partial_{\alpha_j} {\bf{u}}_k$, $i \ne j \ne k \ne i$, lead to same results, respectively.
\end{The}

\begin{center}
\begin{figure}[h]
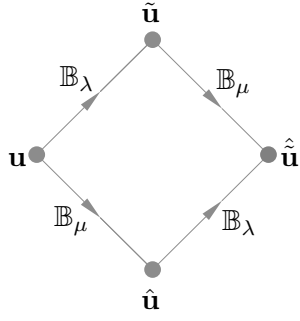

\centertexdraw{ \setunitscale 0.6
\linewd 0.002 \arrowheadtype t:F \arrowheadsize l:0.2 w:0.08
\htext(0 0.5) {\phantom{T}}
\setgray 0.55 \move (1 1) \avec (1.55 .45) \lvec(2 0) \avec (2.6 .6) \lvec (3 1) 
\move (1 1) \avec(1.55 1.55 )\lvec(2 2) \avec(2.6 1.4) \lvec(3 1) 
\move (1 1) \fcir f:0.55 r:0.075 \move (2 0) \fcir f:0.55 r:0.075
\move (2 2) \fcir f:0.55 r:0.075 \move (3 1) \fcir f:0.5 r:0.075
\htext(.75 .9) {$\bf{u}$} \htext(3.1 .9) {$\hat{\tilde{\bf{u}}}$}
\htext(1.9 -.35) {$\hat{\bf{u}}$} \htext(1.9 2.15) {$\tilde{\bf{u}}$}
\htext(1.15 .3) {${\mathbb{B}}_\mu$} \htext(1.2 1.55) {${\mathbb{B}}_\lambda$}
\htext(2.6 .3) {${\mathbb{B}}_\lambda$} \htext(2.55 1.5) {${\mathbb{B}}_\mu$}
}
\caption{Bianchi commuting diagram} \label{fig-Bcd}
\end{figure}
\end{center}

\begin{The}
Let us denote by ${\mathbb{B}}_\lambda({\bf{u}},{\bf{u}}_1,{\bf{u}}_2;\tilde{\bf{u}},\tilde{\bf{u}}_1,\tilde{\bf{u}}_2)$ the algebraic-differential system
\begin{subequations}
\begin{eqnarray}
\frac{\partial \tilde{u}}{\partial \alpha} &=& \frac{\lambda (\lambda u_1-\alpha \tilde{u}) (\lambda \tilde{u} v_1-\alpha u_1 \tilde{v})}{(\lambda^3 - \alpha^3) u u_1 v_1}\frac{\partial u}{\partial \alpha} - \frac{\alpha (\lambda u_1-\alpha \tilde{u}) (\alpha v_1-\lambda \tilde{v})}{(\lambda^3 - \alpha^3) v v_1}\frac{\partial v}{\partial \alpha}\nonumber  \\
&& - n\,\frac{\lambda (u_1 \tilde{v} (\lambda u_1-\alpha \tilde{u}) + v_1 (\alpha u_1^2 - \lambda \tilde{u}^2))}{(\lambda^3-\alpha^3) u_1 v_1} \,,\\
\frac{\partial \tilde{v}}{\partial \alpha} &=& \frac{\alpha (\alpha v_1-\lambda \tilde{v}) (\lambda \tilde{u} v_1-\alpha u_1 \tilde{v})}{(\lambda^3 - \alpha^3) u u_1 v_1}\frac{\partial u}{\partial \alpha} - \frac{(\lambda \tilde{v}-\alpha v_1) (\alpha^2 \tilde{v}-\lambda^2 v_1)}{(\lambda^3 - \alpha^3) v v_1}\frac{\partial v}{\partial \alpha}\nonumber  \\
&& - n\,\frac{\lambda (\tilde{u} v_1 (\alpha v_1-\lambda \tilde{v}) + u_1 (\lambda v_1^2 - \alpha \tilde{v}^2))}{(\lambda^3-\alpha^3) u_1 v_1}\,,\\
\frac{\partial \tilde{u}}{\partial \beta} &=& \frac{\lambda (\lambda u_2-\beta \tilde{u}) (\lambda \tilde{u} v_2-\beta u_2 \tilde{v})}{(\lambda^3 - \beta^3) u u_2 v_2}\frac{\partial u}{\partial \beta} - \frac{\beta (\lambda u_2-\beta \tilde{u}) (\beta v_2-\lambda \tilde{v})}{(\lambda^3 - \beta^3) v v_2}\frac{\partial v}{\partial \beta}\nonumber  \\
&& - m\,\frac{\lambda (u_2 \tilde{v} (\lambda u_2-\beta \tilde{u}) + v_2 (\beta u_2^2 - \lambda \tilde{u}^2))}{(\lambda^3-\beta^3) u_2 v_2} \,,\\
\frac{\partial \tilde{v}}{\partial \beta} &=& \frac{\beta (\beta v_2-\lambda \tilde{v}) (\lambda \tilde{u} v_2-\beta u_2 \tilde{v})}{(\lambda^3 - \beta^3) u u_2 v_2}\frac{\partial u}{\partial \beta} - \frac{(\lambda \tilde{v}-\beta v_2) (\beta^2 \tilde{v}-\lambda^2 v_2)}{(\lambda^3 - \beta^3) v v_2}\frac{\partial v}{\partial \beta}\nonumber  \\
&& - m\,\frac{\lambda (\tilde{u} v_2 (\beta v_2-\lambda \tilde{v}) + u_2 (\lambda v_2^2 - \beta \tilde{v}^2))}{(\lambda^3-\beta^3) u_2 v_2}\,,
\end{eqnarray}
\begin{equation}
\tilde{u}_1 = v\,\frac{\alpha \tilde{u} - \lambda u_1}{\alpha v_1-\lambda \tilde{v}}\,,\quad \tilde{v}_1 = \frac{v}{u} \frac{\alpha u_1 \tilde{v} - \lambda \tilde{u} v_1}{\alpha v_1-\lambda \tilde{v}}\,,\quad \tilde{u}_2 = v\,\frac{\beta \tilde{u} - \lambda u_2}{\beta v_2-\lambda \tilde{v}}\,,\quad \tilde{v}_2 = \frac{v}{u} \frac{\beta u_2 \tilde{v} - \lambda \tilde{u} v_2}{\beta v_2-\lambda \tilde{v}}.
\end{equation}
\end{subequations}
System ${\mathbb{B}}_\lambda({\bf{u}},{\bf{u}}_1,{\bf{u}}_2;\tilde{\bf{u}},\tilde{\bf{u}}_1,\tilde{\bf{u}}_2)$ maps any solution $({\bf{u}},{\bf{u}}_1,{\bf{u}}_2)$ of system $B_{n,m}$ to a new solution $(\tilde{\bf{u}},\tilde{\bf{u}}_1,\tilde{\bf{u}}_2)$ of the same system. In other words, it constitutes an auto-B{\"a}cklund transformation of system (\ref{con-sys-mbsq}).

Moreover, starting with any solution $({\bf{u}},{\bf{u}}_1,{\bf{u}}_2)$ of system $B_{n,m}$ and using the auto-B{\"a}cklund transformations ${\mathbb{B}}_\lambda$ and ${\mathbb{B}}_\mu$ to construct two new solutions $(\tilde{\bf{u}},\tilde{\bf{u}}_1,\tilde{\bf{u}}_2)$ and $(\hat{\bf{u}},\hat{\bf{u}}_1,\hat{\bf{u}}_2)$, respectively, then a third solution $(\hat{\tilde{\bf{u}}},\hat{\tilde{\bf{u}}}_1,\hat{\tilde{\bf{u}}}_2)$ can be constructed algebraically according to Bianchi commuting diagram, cf. Figure \ref{fig-Bcd}, and this solution is given by
\begin{eqnarray*}
&& \hat{\tilde{u}}\,=\,v\,\frac{\lambda \hat{u} - \mu \tilde{u}}{\lambda \tilde{v} - \mu \hat{v}}\,,\qquad \quad \hat{\tilde{v}}\,=\,\frac{v}{u}\,\frac{\lambda \tilde{u} \hat{v} - \mu \hat{u} \tilde{v}}{\lambda \tilde{v} - \mu \hat{v}}\,,\\
&& \hat{\tilde{u}}_1\,=\,v_1\,\frac{\lambda \hat{u}_1 - \mu \tilde{u}_1}{\lambda \tilde{v}_1 - \mu \hat{v}_1}\,,\quad \hat{\tilde{v}}_1\,=\,\frac{v_1}{u_1}\,\frac{\lambda \tilde{u}_1 \hat{v}_1 - \mu \hat{u}_1 \tilde{v}_1}{\lambda \tilde{v}_1 - \mu \hat{v}_1}\,,\\
&& \hat{\tilde{u}}_2\,=\,v_2\,\frac{\lambda \hat{u}_2 - \mu \tilde{u}_2}{\lambda \tilde{v}_2 - \mu \hat{v}_2}\,,\quad \hat{\tilde{v}}_2\,=\,\frac{v_2}{u_2}\,\frac{\lambda \tilde{u}_2 \hat{v}_2 - \mu \hat{u}_2 \tilde{v}_2}{\lambda \tilde{v}_2 - \mu \hat{v}_2}\,,
\end{eqnarray*}
the form of which coincides with the lattice modified Boussinesq system (\ref{mbsq-sys}).
\end{The}

\begin{The} 
The linear system
\begin{subequations}\label{lax-pair-con-sys-mbsq}
\begin{equation}
\partial_\alpha \Psi = K_n({\bf{u}},{\bf{u}}_1,\partial_\alpha {\bf{u}};\alpha) \Psi\,,\quad \partial_\beta \Psi = K_m({\bf{u}},{\bf{u}}_2,\partial_\beta {\bf{u}};\beta) \Psi\,,
\end{equation}
where
\begin{equation}
K_n({\bf{u}},{\bf{u}}_1,\partial_\alpha {\bf{u}};\alpha)  := 
\frac{1}{\alpha^3-\lambda^3} \left(\begin{array}{ccc} 
\frac{\alpha^3  \partial_\alpha v}{v} & \frac{\lambda^2  (n u - \alpha \partial_\alpha u)}{u_1} & \frac{\alpha \lambda  (\alpha v \partial_\alpha u + u (n v - \alpha \partial_\alpha v))}{u v_1}\\
\frac{\alpha \lambda u_1}{u} \frac{n v + \alpha \partial_\alpha v}{v} & \frac{-\alpha^3 \partial_\alpha u}{u} & \frac{\lambda^2 u_1 (\alpha v \partial_\alpha u + u (n v- \alpha \partial_\alpha v)}{u^2 v_1} \\
\frac{\lambda^2 v_1 (n v + \alpha \partial_\alpha v)}{v^2} & \frac{\alpha \lambda v_1 (n u - \alpha \partial_\alpha u)}{u v} & \alpha^3 \left(\frac{\partial_\alpha u}{u}-\frac{\partial_\alpha v}{v} \right)
\end{array} \right)\,,
\end{equation}
\end{subequations}
constitutes a Lax pair of system $B_{n,m}$. In other words, the compatibility condition
$$\partial_\beta K_n({\bf{u}},{\bf{u}}_1,\partial_\alpha {\bf{u}};\alpha) - \partial_\alpha K_m({\bf{u}},{\bf{u}}_2,\partial_\beta {\bf{u}};\beta) + [K_n({\bf{u}},{\bf{u}}_1,\partial_\alpha {\bf{u}};\alpha), K_m({\bf{u}},{\bf{u}}_2,\partial_\beta {\bf{u}};\beta)] =0$$
holds on solutions of system (\ref{con-sys-mbsq}).
\end{The}

\section{Conclusions} \label{sec-Concl}

In this paper we have investigated the symmetry structure and conservation laws for two key examples of higher order integrable discrete systems, 
namely the lattice Boussinesq and the lattice modified Boussinesq system of partial difference equations. A third example in this class is the 
lattice Schwarzian Boussinesq equation, \cite{N2}, which can also be written as a quadrilateral system of three fields, \cite{W}, and for which 
we expect the symmetry analysis to be similar to that of the one performed in Section \ref{sec-Bsq}. Last year, in \cite{H}, a systematic search for integrable 
cases of 2- and 3-field systems of Boussinesq-type form revealed the existence of certain parameter-deformations of the Boussinesq lattice systems, 
and recently in \cite{ZZN} it was shown that these systems arise from a direct linearization scheme similar to the one exploited in \cite{NPCQ} but 
involving a more general dispersion structure. We expect that the deformed Boussinesq systems, including the analogues of the Schwarzian Boussinesq case, 
admit a similar analysis as the one performed in the present paper, but we postpone those generalizations to a future publication. 

Another related aspect, which is currently in preparation, is the existence of Lagrange structures for the Boussinesq systems. The scalar 9-point equation 
(\ref{eq:dbsq}) admits a natural Lagrangian, which was already reported in \cite{NPCQ}, but a Lagrangian for the modified lattice Boussinesq equation 
(\ref{eq:dmbsq}) was not given. In fact, in \cite{LN} it was shown that the entire lattice Gel'fand-Dikii hierarchy, which comprises the lattice potential 
KdV equation as well as the lattice Boussinesq system and associated higher order systems, admits a Lagrangian structure which can be written in 
a surprisingly compact form. Recently, Lagrangians were established for the modified Boussinesq system comprising the fully discrete, semi-discrete systems as 
well as the corresponding generating PDE, \cite{LNX}. These Lagrangian structures allow us to investigate the symmetries and conservation laws for these
integrable systems from the point of view of the relevant Noether theorems. These aspects are part of ongoing research.

\section*{Acknowledgments}

P.X. was supported by the {\emph{Newton International Fellowship}} grant NF082473 entitled
``Symmetries and integrability of lattice equations and related partial differential equations'', which is run by The British Academy, The Royal Academy of Engineering and The Royal Society. F.W.N. is supported by a Royal Society/Leverhulme Trust Senior Research Fellowship (2011/12).


\begin{thebibliography}{99}

\bibitem{ABS} Adler V.E., Bobenko A.I. and Suris Yu.B. (2003) Classification of integrable equations on quad-graphs. The consistency approach {\em Comm. Math. Phys.} {\bf{233}} 513--543

\bibitem{DJM} Date E., Jimbo M. and Miwa T. (1983) Method for generating discrete soliton equations III, {\em J. Phys.  Soc. Japan} 
{\bf 52} 388--393. 

\bibitem{FF} 
Finkel F. and Fokas A.S. (2002) On the construction of evolution equations admitting a master symmetry {\em{Phys. Lett. A}} {\bf{293}} 36--44

\bibitem{H}
Hietarinta J. (2011) Boussinesq-like multi-component lattice equations and multi-dimensional consistency 
{\em{J. Phys. A: Math. Theor.}} {\bf 44}, no. 16, 165204 (22 pp).

\bibitem{LN}
Lobb, S., and F. W. Nijhoff (2010) Lagrangian multiform structure for the lattice Gel'fand-Dikii hierarchy 
{\em{J. Phys. A: Math. Theor.}} {\bf 43} 072003.

\bibitem{LNX} 
Lobb S., Nijhoff F.W. and Xenitidis P., in preparation.  

\bibitem{M} Mikhailov A.V. (ed.) (2009) {\em Integrability} Lecture Notes in Physics {\bf{767}}, Springer (2009)

\bibitem{MWX1} Mikhailov A. V., Wang J. P.  and Xenitidis P. (2011) Recursion operators, conservation laws and integrability conditions for difference equations {\emph{Theor. Math. Phys.}} {\bf{167}} 421--443

\bibitem{N2}
Nijhoff F.W. (1996) On some ``Schwarzian Equations'' and their Discrete Analogues. in: Eds. A.S. Fokas and I.M. Gel'fand,
{\it Algebraic Aspects of Integrable Systems: In memory of Irene Dorfman}, (Birkh\"{a}user Verlag), 237--60.

\bibitem{N1} Nijhoff F. W. (1999) Discrete Painlev{\`e} Equations and Symmetry Reduction on the Lattice, {\sl{in ``Discrete Integrable Geometry and Physics''}}, eds. A. I. Bobenko \& R. Seiler, Oxford Lecture Series in Mathematics and Its Applications, OUP

\bibitem{NJH} Nijhoff F. W., Hone A. and Joshi N. (2000) On a Schwarzian PDE associated with the KdV hierarchy {\emph{Phys. Lett. A}} {\bf{267}} 147--156

\bibitem{NPCQ} Nijhoff F. W., Papageorgiou V. G., Capel H. W. and Quispel G. R. W. (1992) The lattice Gel'fand-Dikii hierarchy {\emph{Inverse Problems}} {\bf{8}} 597--651

\bibitem{O} Olver P.J. (1993) {\emph{Applications of {L}ie groups to Differential Equations}} Graduate Texts in Mathematics {\bf{107}} Springer Verlag

\bibitem{TN2} Tongas A. and Nijhoff F. W. (2005) The Boussinesq integrable system: Compatible lattice and continuum structures {\emph{Glasgow Math. J.}} {\bf{47A}} 205--219

\bibitem{TN1} Tongas A. and Nijhoff F. W. (2006) A discrete Garnier type system from symmetry reduction on the lattice {\emph{J. Phys. A: Math. Gen.}} {\bf{39}} 12191--12202

\bibitem{TX} Tsoubelis D. and Xenitidis P. (2009) Continuous symmetric reductions of the Adler-Bobenko-Suris equations {\em{J. Phys. A: Math. Theor.}} {\bf 42} 165203 (29pp)

\bibitem{W} Walker A. J. (2001) {\emph{Similarity reductions and integrable lattice equations}} Ph.D. thesis, University of Leeds

\bibitem{X} Xenitidis P. (2011) Symmetries and conservation laws of the ABS equations and corresponding differential-difference equations of Volterra type {\em{J. Phys. A: Math. Theor.}} {\bf{44}} 435201 (22pp)

\bibitem{XNL} Xenitidis P., Nijhoff F. W. and Lobb S. (2011) On the Lagrangian formulation of multidimensionally consistent systems {\emph{Proc. R. Soc. A}} {\bf{467}} 3295-3317 

\bibitem{Z} Zakharov V.E. (ed.) (1991) {\em What is integrability?} Springer Verlag, Berlin

\bibitem{ZZN} 
Zhang D.-J., Zhao S.-L. and Nijhoff F.W. (2011)  Direct Linearization of extended lattice BSQ systems {\it{arxiv:1112.0525}} (submitted to {\em{Studies in Applied Mathematics}}) 

\end{thebibliography}
\end{document}